# Universal Relation between Nusselt Number and Bejan Number in Natural Convection

Takuya Masuda [*]

Department of Integrated Engineering, National Institute of Technology, Yonago College,

4448 Hikona-cho, Yonago, Tottori 683-8502, Japan

Toshio Tagawa

Department of Aeronautics and Astronautics, Tokyo Metropolitan University,

6-6 Asahigaoka, Hino, Tokyo 191-0065, Japan

[*] Contact author: masuda@yonago-k.ac.jp

We propose a universal scaling law linking the Nusselt number ($Nu$) and the Bejan number ($Be$) in natural convection. Using entropy generation analysis and boundary-layer scaling, we demonstrate that $Be^{-1} - 1 = a\, Nu^b$ emerges independently of geometry and boundary conditions when transport is governed by a single control parameter. The relation is validated using a canonical square cavity. This result establishes a direct connection between heat transfer and thermodynamic irreversibility, revealing a fundamental constraint in convective transport.

Natural convection is one of the most fundamental transport mechanisms in fluid mechanics, governed primarily by buoyancy forces. Its heat transfer performance is commonly quantified by the Nusselt number $Nu$, which is known to obey power-law scaling with respect to the Rayleigh number $Ra$: $Nu \sim Ra^n$ [1–3]. This scaling behavior has been extensively confirmed across a wide range of configurations, including Rayleigh–Bénard convection [1,2], differentially heated cavities [4,5], and horizontal convection [6]. On the other hand, thermodynamic irreversibility in convective systems is typically evaluated through entropy generation analysis [7–13]. The Bejan number $Be$, defined as the ratio of entropy generation due to heat conduction to the total entropy generation, provides a measure of the relative importance of thermal and viscous dissipation mechanisms. Despite the extensive use of both $Nu$ and $Be$, these two quantities have traditionally been treated independently. In particular, while numerous studies have examined their dependence on governing parameters such as $Ra$ [14–21], a direct functional relationship between them has not been explicitly established.

Recently, a three-dimensional numerical study of thermomagnetic convection under a quadrupole magnetic field demonstrated that the following relation holds [22]:

$$Be^{-1} - 1 = a\, Nu^b, \tag{1}$$

where $a$ and $b$ are constants. Importantly, the derivation of this relation does not explicitly involve geometric parameters or boundary conditions, but instead relies on the scaling structure of entropy generation. This observation suggests that the above relation may represent a universal scaling law, applicable to any natural convection system in which $Nu$ follows a power-law dependence on a single control parameter such as $Ra$.

The objective of this Letter is to formalize this idea and demonstrate that the relation of Eq. (1) is a general consequence of scaling laws in natural convection, independent of geometry and boundary conditions. Furthermore, this universality is demonstrated using a canonical benchmark problem: natural convection in a side-heated square cavity [4,5].

The non-dimensional governing equations are shown as follows:

$$\nabla \cdot \mathbf{U} = 0, \tag{2}$$

$$\frac{D\mathbf{U}}{DT} = -\nabla P + Pr\nabla^2 \mathbf{U} - PrRa\Theta \mathbf{e}_y, \tag{3}$$

$$\frac{D\Theta}{DT} = \nabla^2 \Theta. \tag{4}$$

The total entropy generation within the domain is calculated as the sum of the thermal and viscous contributions:

$$s = s_\theta + s_\psi. \tag{5}$$

The dimensionless local entropy generation rates due to heat transfer and fluid friction, denoted by $S_{\Theta,loc}$ and $S_{\Psi,loc}$, are defined as follows:

$$S_{\Theta,loc} = |\nabla\Theta|^2, \tag{6}$$

$$S_{\Psi,loc} = 2\varphi \mathbf{D} : \mathbf{D}, \tag{7}$$

where

$$\mathbf{D} = \frac{1}{2}(\nabla\mathbf{U} + (\nabla\mathbf{U})^T). \tag{8}$$

The irreversibility distribution ratio is taken as $\varphi = 10^{-4}$, following previous studies [9,11]. The total entropy generation rates over the entire domain are obtained by integrating the local contributions. The Bejan number is defined as

$$Be = \frac{s_\theta}{s_\theta + s_\psi}. \tag{9}$$

Thus,

$$\frac{1}{Be} - 1 = \frac{s_\psi}{s_\theta}. \tag{10}$$

This expression shows that $Be^{-1} - 1$ directly represents the ratio of viscous to thermal irreversibility.

The scaling behavior of the Nusselt number and entropy generation can be rationalized using boundary-layer arguments. In natural convection, the thermal boundary-layer thickness $\delta_\theta$ follows a power-law dependence on the governing parameter, reflecting the balance between buoyancy-driven flow and diffusive transport. Since the Prandtl number is fixed in the present study, the ratio between the momentum boundary-layer thickness $\delta_u$ and $\delta_\theta$ remains approximately constant. The Nusselt number scales as $Nu \sim l_c/\delta_\theta$. Therefore,

$$Nu \sim Ra^n. \tag{11}$$

Because the temperature gradient in the boundary layer scales as $|\nabla\theta| \sim \delta_\theta^{-1}$, the thermal entropy generation, which is dominated by the boundary-layer contribution, follows

$$s_\theta \sim Ra^p. \tag{12}$$

Similarly, viscous entropy generation is governed by near-wall shear layers, where the velocity gradient scales as $|\nabla\mathbf{u}| \sim u_c/\delta_u$. Consequently,

$$s_\psi \sim Ra^q. \tag{13}$$

Therefore, $Be^{-1} - 1 = a\, Nu^b$ holds. This derivation does not include any geometric or boundary-condition-specific parameters, implying general applicability.

To illustrate the universality of the proposed relation, we consider natural convection in a square cavity with differentially heated vertical walls—a standard benchmark problem widely used for CFD validation. This configuration has been extensively studied since the pioneering work of De Vahl Davis, and its flow and heat transfer characteristics are well established. The numerical setup consists of a square cavity filled with air ($Pr = 0.71$). The left vertical wall is maintained at a constant high temperature, while the right vertical wall is kept at a constant low temperature. The horizontal walls are assumed to be adiabatic. The governing equations are solved using a second-order accurate central difference scheme with finite-volume method, and steady-state solutions are obtained over the Rayleigh number range $10^3 \leq Ra \leq 10^7$.

The result is shown on Fig. 1. Consistent with classical studies, the Nusselt number follows

$Nu \sim Ra^n$. Entropy generation is computed by integrating thermal and viscous contributions over the domain. The resulting data show that $Be^{-1} - 1 = a\,Nu^b$ holds with high accuracy across the entire range of $Ra$.

Despite the strong dependence of flow structure on $Ra$ (transition from conduction-dominated to convection-dominated regimes), the $Nu$–$Be$ relation remains unchanged. This confirms that the relation reflects a fundamental balance between heat transport and irreversibility, rather than details of flow structure.

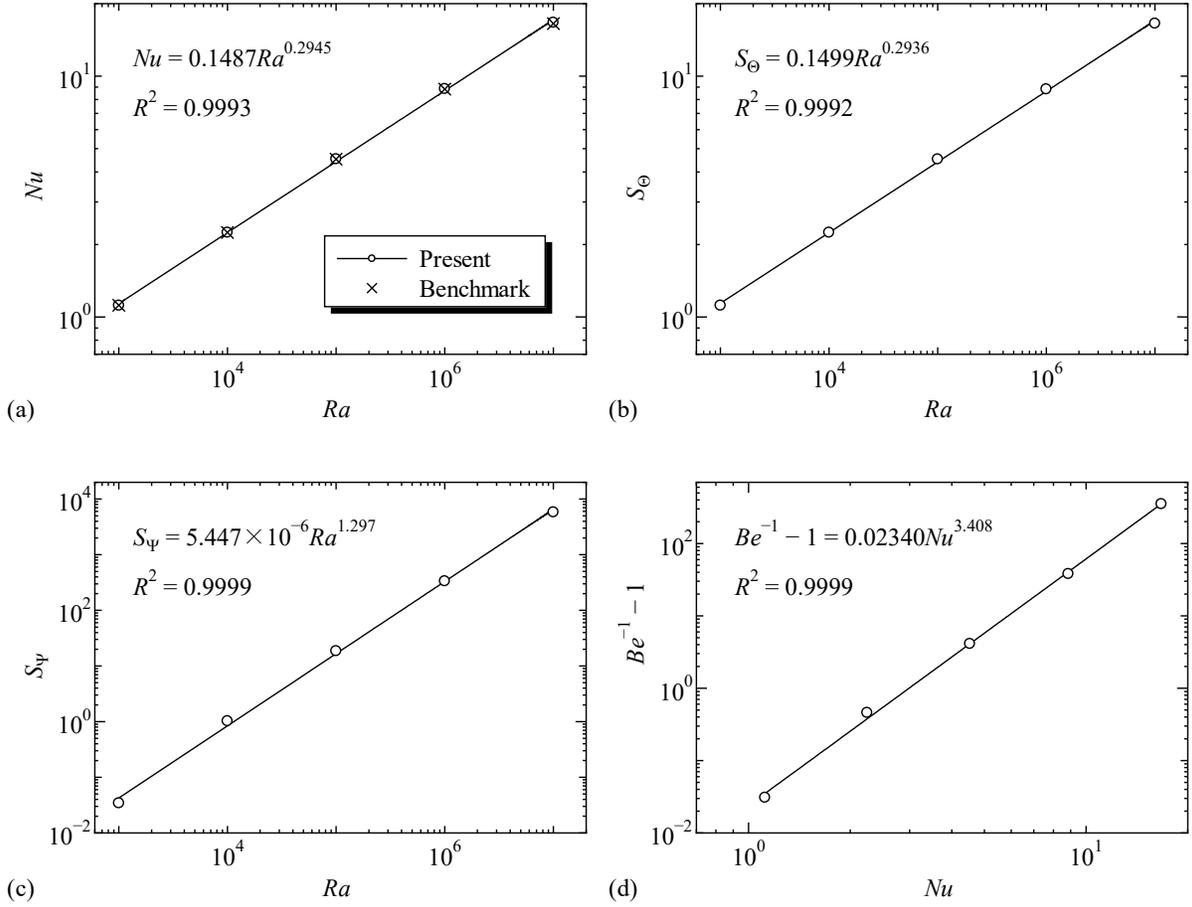

FIG. 1. (a) Average Nusselt number $Nu$, (b) dimensionless total entropy generation due to heat conduction $S_\Theta$, and (c) dimensionless total entropy generation due to viscous dissipation $S_\Psi$ as functions of the Rayleigh number $Ra$, together with (d) the distribution of $Be^{-1} - 1$ as a function of $Nu$. Here, $R^2$ denotes the coefficient of determination. Reference values of $Nu$ are taken from de Vahl Davis [4] for $Ra = 10^3$–$10^5$ and from Le Quéré [5] for $Ra = 10^6$–$10^7$.